# Time-resolved optical assessment of exciton formation in mixed two-dimensional perovskite films

Zheng Zhang[1,2,3], Jianan Wang[1,3], Yijie Shi[1,2], Xi Wang[1,2], Zhong Wang[1,2], Xiangyu Zhu[1,2], Chunlong Hu[1,2], Zonghao Liu[1], Wei Chen[1], Wenxi Liang[1,2*]

[1]Wuhan National Laboratory for Optoelectronics, Huazhong University of Science and Technology, Wuhan 430074, China

[2]Advanced Biomedical Imaging Facility, Huazhong University of Science and Technology, Wuhan 430074, China

[3]These authors contributed equally: Zheng Zhang, Jianan Wang

* Email: wxliang@hust.edu.cn

**Abstract:**

The formation and relaxation of excitons lay significant impacts on processes of carrier transport and recombination in low-dimensional semiconductors, in which the excitonic properties are prominent due to strong confinement effects. These dynamics are subjected to the binding energy which sets the barrier for initiation of the transition. Here, we report the observation of exciton formation from the cooled band-edge electrons and holes in the mixed two-dimensional hybrid organic-inorganic perovskites $(F\text{-}PEA)_2(MA)_{n-1}Pb_nI_{3n+1}$ using femtosecond transient absorption spectroscopy. By monitoring the changes of bleach signal following the variance of photon energy required for the excitations of different electronic states, we extract the values of exciton binding energy and the occupancies of carriers of free and bound states for each two-dimensional phase. We also observe the persistence of the sub-picosecond durations of exciton formation, and the enhanced rate of carrier transfer by one order of magnitude from the two-dimensional phases to the three-dimensional phase, with injected carrier density above the Mott criterion. These results offer us a new approach to evaluate the exciton binding energy through measurements of transient absorption spectroscopy, and insights probably helpful for applications.

Strongly bound excitons form stably in low-dimensional semiconductors at room temperature[1,2], due to the enhanced coulombic interactions by spatial confinement and weakened dielectric screening. Exciton formations take place either by direct excitations or the coupling of free band-edge electrons and holes[3], generated from the relaxation of excited carriers by absorption of photon energy ($E_x$) above the bandgap ($E_g$). The associated interaction, time span, and pathway in these processes determine fundamental dynamics including carrier transfer, energy relaxation, and transitions between the free and bound states[4-7], as well contribute to optoelectronic performances in various aspects. For example, the bound excitonic states hamper the separation of electron-hole pairs, thereby lowering the efficiency of carrier collection in photovoltaic devices[8]; but the strongly bound excitons provide higher recombination rates for the precedence over defect trapping, thereby increasing the luminescence efficiency in light-emitting devices[9].

The exciton binding energy ($E_b$) is a determinant of exciton properties, critical to the strength bonding electron-hole pairs and hence the stability of excitons. Fitting absorption spectra by the Elliott formula, and temperature-dependent photoluminescence (PL) spectra by the Arrhenius formula are currently two main approaches of evaluating $E_b$. However, the former suffers from the large number of parameters involved, while the latter suffers from the concurrence of defect-assisted recombination, Auger recombination, and other temperature-dependent processes[10]. Besides the issue of evaluating $E_b$, the mutual conversions between free carriers and excitons remain an open question, especially under conditions that excitons are believed to ionize into an electron-hole plasma when the amount of photoinjected carriers reaches the Mott density[11]. Suzuki *et al*. found that the proportion of free carriers in silicon increases following the increase of carrier density, contradicting to that predicted by the Saha equation, which are experimentally testified[12,13]. Mahan excitons are observed in perovskites under the condition of Mott criterion[13], in which the population of free carriers is supposed to reach unity. One practical challenge hindering the understanding of exciton properties and the associated dynamics lays in experimentally distinguishing the signals from excitons and free carriers, as these two species of quasiparticle usually populate in energy states with close values within a sub-picosecond time window, the span for the cool down of hot carriers and the formation of excitons[3].

Two-dimensional hybrid organic-inorganic perovskites (2D HOIPs) possess $E_b$ up to hundreds of meV due to the large discrepancy of dielectric permittivity between the organic and inorganic layers. The mix of different phases of 2D HOIPs with distinctive $E_b$ in each phase makes possible the simultaneous examination for different excitonic states in one sample, providing us an excellent platform for investigations on exciton properties. The mixed 2D HOIPs hold advantages from both the 2D component, e.g., tunable bandgap, high stability, and so on, and the three-dimensional (3D) component, e.g., large absorption coefficient and facile solution processability[9,14-18], becoming a hot topic in applications of light harvesting and emitting[9,19]. While 2D HOIPs feature a heterostructure with the inorganic octahedra enveloped by two organic molecules, the mixed 2D HOIPs are naturally constructed with heterojunctions. The former yields an $E_b$ much higher than $k_0T$, resulting in the dominant role of excitons at room temperature; the latter forms a type II band alignment of multi-quantum well structure with staggered bandgaps[20,21], resulting in rapid transfers of electrons and holes preferentially to different phases[22] and the formation of long-lived electron-hole separation states, subsequently an improved extraction efficiency for solar cells[20,21]. Plenty of works on carrier and energy transfers in mixed 2D HOIPs have been reported (see ref. 23, 24 and references therein), but the impacts of free carrier and exciton on processes governing carrier dynamics are not yet elucidated. For example, although the Saha equation predicts that free carriers dominate over excitons with injected carrier density of $10^{10}$ cm$^{-2}$ in 2D HOIPs with an $E_b$ of ~200 meV[25,26], contradicting results are reported that excitons constitute the primary specie[5,8].

In this work we assess the exciton formation during carrier transfers in the mixed 2D HOIPs (F-PEA)$_2$(MA)$_{n-1}$Pb$_n$I$_{3n+1}$ using femtosecond transient absorption (TA) spectroscopy. The observed sub-picosecond decays of bleach signal evolve following the variation of $E_x$, enabling us to identify the formation and dissociation of excitons, and to precisely extract the $E_b$ values for different phases. We also observe the coexistence of free carriers and excitons when the injected carriers exceed the Mott density, resembling the Mahan excitons. To the best of our knowledge, here the assessment presents the first quantitatively evaluation of exciton properties through the well-adopted TA measurements, establishing a new approach to disentangle the kinetics of excitons from those of free carriers.

**Results**

The mixed (F-PEA)$_2$(MA)$_{n-1}$Pb$_n$I$_{3n+1}$ films were synthesized by spin-coating (see Methods), with two thicknesses of 448 and 295 nm for validating the effect of distance between donors and acceptors during the possible energy transfer. The scanning electron microscopy images show less pores on the surface of the thick sample compared to the thin one (see Supplementary Fig. S1), which are attributed to a lower surface defect density. Fig. 1a presents the steady-state absorption and PL spectra measured by illuminations from the front and the back (with a glass substrate) of sample. Both absorption spectra peak at 558, 605, and 760 nm for the n=2, n=3 (referred to as n2 and n3 hereafter, respectively), and 3D (n=∞) phases, respectively. In contrast, the PL spectrum by front excitation shows only a luminescence peak at 778 nm for the 3D phase, while the one by back excitation shows peaks at 571 and 618 nm for n2 and n3 phases, respectively, along with the peak for the 3D phase. The variance in the PL spectra upon two excitation directions suggests a distribution of different perovskite phases along the direction normal to the substrate, in which the phases with low n values grow close to the substrate while the 3D phase prefers the front side.

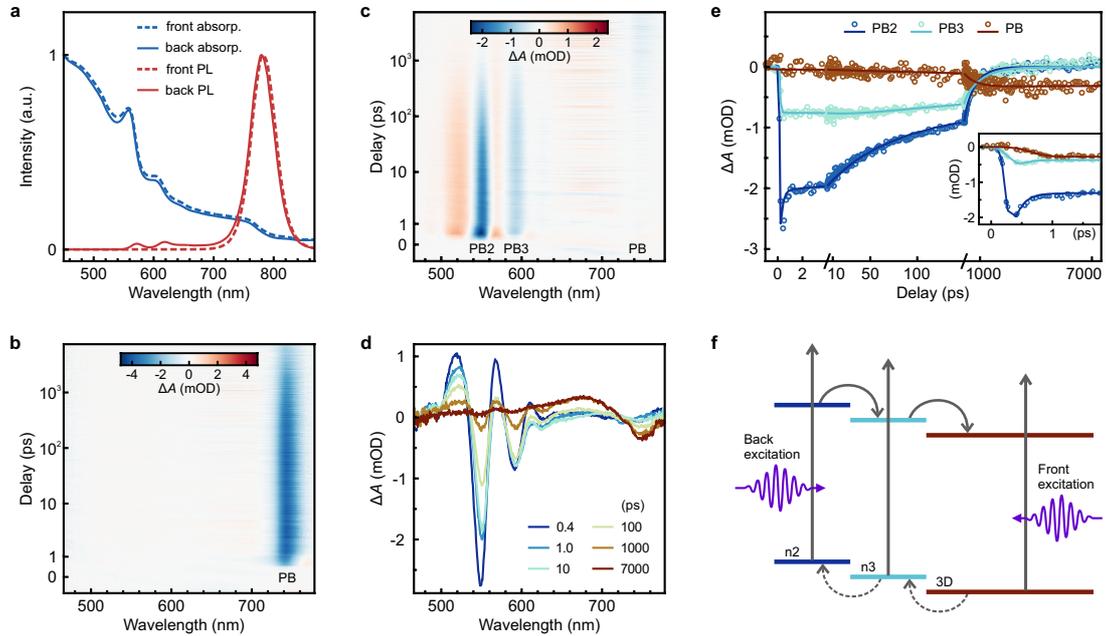

**Figure 1 Optical characterizations and carrier transfers in the mixed (F-PEA)$_2$(MA)$_{n-1}$Pb$_n$I$_{3n+1}$. a**, Steady-state absorption spectra (blue) and PL spectra (red) measured with front (dashed lines) and back (solid lines) excitations. The PL spectra were excited by 405 nm wavelength. Pseudocolor contour plots of TA measurement by front (**b**) and back (**c**) excitations with 400 nm and fluence of 2.6 μJ/cm$^2$, showing only a long lasting PB signal in **b**, and bleach signals of PB2, PB3, and PB, along with a very weak signal peaking at 625 nm for the phase of n=4 (see text) in **c**. **d**, Spectra extracted at selected delay times from **c**. **e**, Kinetic traces of PB2, PB3, and PB extracted from **c**, fitted with multiexponential function (fitting parameters tabulated in Supplementary Table S1). The breaks on the x-axis represent the delay times of 4 and 150 ps, respectively. Inset, a zoomed-in view of the kinetics measured with more sampling points in the early stage after excitation, addressing the sub-picosecond decays of PB2

and PB3. **f**. Schematic diagram of staggered band alignment and carrier transfers in the mixed 2D HOIPs. Solid and dashed arrows denote the transitions of electrons and holes, respectively.

**Carrier transfer in different phases**

We first examine the sample of 448 nm through TA measurements (see Methods) with $E_x$ well above the bandgap. Shown in Fig. 1b is the pseudocolor contour plot by front excitation, showing only the bleach peak at 745 nm for the 3D phase (PB), no discernable features from other phases (see extracted spectra in Supplementary Fig. S3). Shown in Fig. 1c is the pseudocolor plot by back excitation, showing bleach peaks at 551 nm and 592 nm, for n2 (PB2) and n3 (PB3) phases, respectively, and a weak PB at 745 nm. The variance of TA spectral profiles under two excitation directions resembles the results of PL measurement, confirming the distribution of different phases in the sample.

In the result of back excitation, the amplitudes of PB2 and PB3 rise, then reach their maximum during the early 0.4 ps after photoexcitation, as shown in Fig. 1d, reflecting the relaxation of excited hot carriers. In 0.4–1 ps, PB2 rapidly decreases, accompanied with a redshift of peak position, while PB3 shows a slight decrease with less redshift. In 1–100 ps, PB2 shows a slow decay with barely discernable peak shift, while PB3 remains nearly unchanged in amplitude and peak position. These decay processes are correlated to the exciton formation and dissociation, which are discussed in next sections. The PB2 keeps decreasing in the following duration, while PB3 decreases slowly in a duration up to hundreds of picoseconds until eventually decreases concertedly with PB2, as shown in Fig. 1e. The PB evolves in sharp contrast to that by front excitation. The results measured with two excitation configurations outline the scenario of carrier transfers in different phases, see Supplementary Discussion 1 for details, and Supplementary Discussion 2 for the verification of staggered band alignment in the sample. It is noteworthy that the lifetime of PB by back excitation is about 3 times that by front excitation, which is attributed to the electron-hole separation states created after carrier transfer[20]. We conclude the band alignment and the carrier transfer pathway as schematically depicted in Fig. 1f, in which excited electrons flow from the phase with low n value to that with high n value, while excited holes flow in an inverse direction with lower efficiency due to larger effective mass.

**Formation of excitons in the 2D phases**

The sub-picosecond decays of PB2 and PB3, better shown in the inset of Fig. 1e, are commonly interpreted as energy transfer[8,27], defect associated capture[28], or exciton formation[29]. We rule out the processes of energy transfer, defect capturing, and other candidates by experimental observations comparing to results reported in literature, see Supplementary Discussion 3 for details. Both decays strikingly diminish when the $E_x$ reduces, as shown in Fig. 2a and 2b. The one of PB2 diminishes with redshift upon $E_x$ ranging in 400–480 nm, vanishes upon 500–525 nm, then re-appears with blueshift upon 550 nm, until the PB2 signal is totally gone when $E_x$ beyond 560 nm. The same fashion with reduced characteristic $E_x$ applies to that of PB3. Multiexponential fits to the kinetics reveal the lifetimes of the decays increase during the diminishment, with decreasing weights until their vanishing, as depicted in Fig. 2c and 2d. We can rule out the relaxation of hot carriers for these processes because the former is reflected in the rise of bleach signal rather than decay[30], with an increasing rate following the decrease of $E_x$. Fig. 2e summarizes the fitted lifetimes and weights for all measured $E_x$, clearly illustrating the vanishing of the decays upon certain $E_x$ values. The significant changes of the decays by different $E_x$ suggest their origins from variations in energy space.

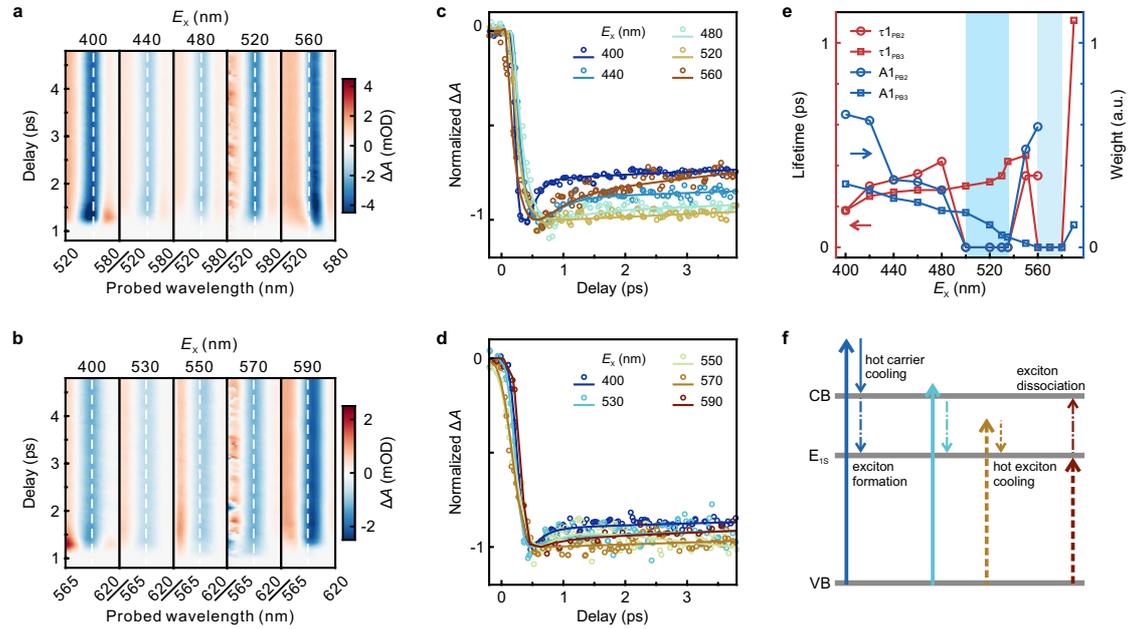

**Figure 2 Exciton formations reflected in TA measurements with varied $E_x$. a** and **b**, Pseudocolor contour plots of the sub-picosecond decays with representative $E_x$, for PB2 and PB3, respectively. The dashed white lines located at 551 nm (**a**) and 592 nm (**b**) are the guides to the eye. Note that the noisy results measured upon excitations of 520 nm (**a**) and 570 nm (**b**) are due to the pump wavelengths close to the probed signals. **c** and **d**, Kinetic traces extracted

from **a** and **b** with multiexponential fits (fitting parameters tabulated in Supplementary Table S4 and S5), for PB2 and PB3, respectively. **e**, Trends of fitted lifetime and weight of the sub-picosecond decays. Blue shaded areas highlight the values of $E_x$ for the vanishing of sub-picosecond decay. **f**, Schematic diagram of transitions of free and bound carriers in the sub-picosecond duration, including population/relaxation (upward/downward solid arrows) of free carriers, and formations/dissociation (downward/upward dash-dot arrows) and populations/relaxation (upward/downward dashed arrows) of excitons, respectively, upon excitations with different $E_x$ tagged by colors. CB, $E_{1s}$, and VB denote the conduction band minimum, the ground state of exciton, and the valence band maxima, respectively.

Given the fact that excitons are widely considered the predominant carriers in 2D perovskites[31,32], which feature larger $E_b$ compared to the 3D counterpart[1,33], the observed occurrences of sub-picosecond decays only in PB2 and PB3 but not in PB lead us to speculate the correlation between the origins of these processes and the $E_x$ of different phases in the sample. Combining the reported hundreds of femtoseconds for exciton formation in 2D perovskites[3], and the contribution from free carriers more than excitons to the amplitude of bleach signal[29,34], we believe that these rapid decay processes may be associated to the formation and dissociation of excitons. In this picture of exciton formation, the relaxation of hot carriers gives rise to PB2 and PB3; then the free electron-hole pairs at band edges release energy to bind as excitons[35], assisted by emissions of optical phonons, resulting in the sub-picosecond decays. The available phonon modes are suppressed as $E_x$ decreases towards $E_g$, resulting in reduced available pathways for the relaxation of free carriers[36], hence the reduced rate of exciton formation, i.e., the observed increasing lifetimes. Conversely, a larger proportion of excited carriers relax to the band edges in the form of free carriers[33] with the increase of $E_x$, resulting in more excitons formed, hence the observed increasing weights.

The relaxations and transitions of carriers between the free and bound states are schematically summarized in Fig. 2f. Hot excitons, instead of hot free-carriers, are initially populated with $E_x$ lower than $E_g$, then cool down to fill the excitonic band, hence the absences of sub-picosecond decay due to no exciton formation from band-edge free carriers. When the $E_x$ further decreases to the ground state of excitonic band ($E_{1s}$), initially populated excitons dissociate to free electron-hole pairs, resulting in the sub-picosecond decays in PB2 and PB3 with a blue shift. The shifts in energy space of PB2 and PB3 after the completeness of sub-picosecond decay corroborate the scenario of exciton dynamics. We observe that PB2 at 1ps red-shifts from 551 nm (400 nm $E_x$) to 557 nm (560 nm $E_x$), see Fig. 2a, and Supplementary Fig. S10 for better illustration.

In the latter case of excitonic excitation, the dominant contributors to bleach signal at 1 ps are not-yet-dissociated excitons, resulting in the lowered energy compared to those in cases of above-bandgap excitation, in which more free carriers with hundreds of meV extending over excitons in 2D perovskites[1], constitute the bleach signals. PB3 shows a same trend but with smaller energy shift, see Fig. 2b and Supplementary Fig. S11, attributed to a smaller energy discrepancy between free and bound states due to the smaller $E_b$ of n3 phase. The sub-picosecond decays are followed by rather slow decays with lifetimes of tens of picoseconds, also showing distinctive energy shifts. The PB2 excited by 400 nm barely red-shifts in the duration of 1–100 ps, while the one excited by 560 nm apparently blue-shifts from 557 nm to 554 nm, as shown in Fig. 3a and 3b, suggesting a two-step dissociation of excitons, with the sub-picosecond process correlated with exciton dephasing[37] and the slow process with many-body interactions, see Supplementary Discussion 4 for more details.

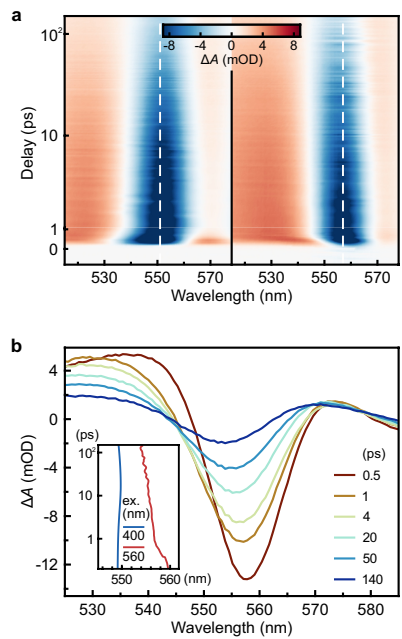

**Figure 3 Exciton dissociation reflected in TA measurements. a,** Pseudocolor contour plots of the slow decays of PB2 upon back excitations of 400 nm (left) and 560 nm (right). The dashed white lines located at 551 nm (left) and 557 nm (right) are the guides to the eye. **b,** Spectra of selected delay times extracted from the right panel of **a**, showing an apparent blue-shift. Inset, comparison of peak positions extracted by Gaussian fits to the spectral profiles following delay times of PB2 excited by 400 nm and 560 nm, and fluences of 24.5 μJ/cm$^2$ and 28.8 μJ/cm$^2$, respectively.

Based on this scenario of exciton dynamics, we are able to evaluate $E_b$ using the $E_x$ values required for the signal of exciton formation, calculated as the difference between $E_g$ and $E_{1s}$, i.e., $E_b = E_g - E_{1s}$, see Supplementary Discussion 5 for more details. In n2 phase, the $E_g$ of 500 nm and $E_{1s}$ of 557 nm of yield an $E_b$ of 253 meV. An $E_b$ of 133 meV for n3 phase is obtained by the same method. These values are consistent with the reported results by density functional theory calculations and fitting of steady-state absorption spectra[1,7], 250/170 and 180/120 meV for n2 and n3 phases, respectively. We can also estimate the occupation proportion of free carriers at thermal equilibrium by substituting the obtained $E_b$ values into the Saha equation, as larger $E_b$ gives rise to more contribution from exciton formation to the decay of bleach signal, see Supplementary Discussion 6 for more details. At last, the contributions from free carriers and excitons to the bleach signal can be quantified. Assuming carriers generated by 400 nm excitation are all free carriers, given the initial temperature of carriers can reach thousands of Kelvins upon above-bandgap excitation[38], we calculated that the contribution of free carriers is 1.33 and 1.25 times that of excitons in n2 and n3 phases, respectively, see Supplementary Discussion 7 for more details.

**Exciton formation with injected carrier density above the Mott criterion**

It is commonly believed that excitons totally ionize with injected carriers of the Mott density[39,40], but the existence of excitonic states proposed by Mahan[41] is confirmed by Palmieri *et al.*[13] In the regime of linear absorption (see the inset in Fig. 4a), the sub-picosecond decays of PB2 remain unchanged in both rate and weight, as shown in Fig. 4a and Supplementary Table S9, indicating the proportions of free carriers and excitons remaining unchanged at the quasi-equilibrium, and the negligible impact of weak screening. We increased the carrier density injected into the sample higher than the Mott criterions, see Supplementary Discussion 8 for details, finding that both the rate and weight decrease, as shown in Fig. 4b and Supplementary Table S9, in the regime of saturating absorption (see the inset in Fig. 4b). The weakening process of exciton formation may be interpreted with the filling of low energy states resulting in the strengthened screening, hence the reduction of $E_b$ and higher proportion of free carriers. Additionally, the high injected carrier density possibly induces a strong bandgap renormalization also enhancing the screening[42], hence the observed changes. The fitted lifetimes and weights of the sub-picosecond decays of PB2 for all measured excitation fluences are depicted in Fig. 4c, illustrating the impacts of screening on the exciton formation in a much larger regime of nonlinear absorption.

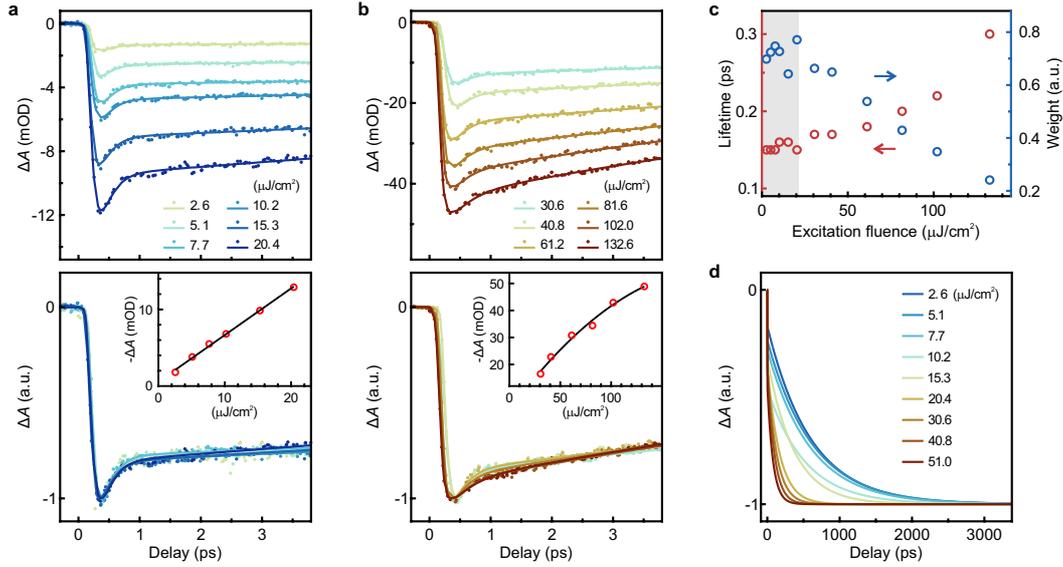

**Figure 4 Exciton formations and carrier transfers reflected in TA measurements with injected carrier densities approaching the Mott criterion**. **a** and **b,** Kinetic traces (upper panels) and the normalized traces (lower panels) of PB2 with multiexponential fits, upon back excitations of 400 nm and fluences of 2.6–20.4 μJ/cm$^2$ and 30.6–132.6 μJ/cm$^2$, respectively. Insets, linear (**a**) and saturating (**b**) trends of the PB2 amplitudes following the increase of excitation fluence. The black lines are the guides to the eye. **c,** Fitted lifetimes and weights of the sub-picosecond decays of PB2 for all measured excitation fluences. The gray shaded area highlights the linear absorption regime. **d,** Kinetics of PB signal following the increase of excitation fluence, depicted with the fitted traces (fitting parameters tabulated in Supplementary Table S10) for clearance, normalized. See Supplementary Fig. S15 for the measured kinetics.

The kinetic traces of PB3 evolve similarly with changed features corroborating the picture of Mahan excitons, see Supplementary Fig. S14 for details. The persisting processes of exciton formation indicate that, with injected carriers notably exceeding the Mott density, carriers in the mixed 2D HOIPs are not totally free at the quasi-equilibrium before interband relaxation, but rather a small fraction of Mahan excitons. The screening also lay impacts on the processes of carrier transfer, due to the better mobility of free carriers over excitons. We observe the rise times of PB significantly shortened following the increase of excitation fluence, by an order of magnitude at best, as depicted in Fig. 4d.

**Conclusions**

Our observations of exciton formations in the mixed 2D HOIPs demonstrate the controllable ratio of populations in free and bound states, to some extent, by varying the $E_x$ and fluence. The complexity of bleach signal decay convoluted with multiple species of carrier dynamics, lays an obstacle for evaluating the mobilities of free carriers and excitons in heterostructures. Here we are able to extract the transfer rate of free carriers from the 2D phases into the 3D phase, by properly chosen excitation conditions for suppressing the proportion of excitons formed by free carriers. The lifetime of free carriers is longer than that of excitons by about one order of magnitude, see Supplementary Figure S16, benefiting carrier extractions; but a fair portion of unbound carriers in 2D HOIPs are involved in the formation of excitons, especially in cases of low injected carrier density. Given the carrier extraction time on the order of nanosecond[43], the inefficient utilization of photogenerated carriers is considered a key factor accounting for the lower efficiency of solar cell by 2D perovskites compared to their 3D counterparts[19,23,44]. The ever-persisted exciton formations drive one to seek for strategies either to make use of these bound-state carriers, or to exploit the tunable transfer rate across heterojunctions, for implements of mixed 2D HOIPs.

Monitoring the changes of bleach signal under varied excitation conditions presented here offers us the possibility of establishing a new approach, to quantitatively assess the formation of excitons and subsequent evolutions. The extraction and evaluation of energy, time span, and pathway of transition between different states associated to these processes, build us the base for further insights into the characteristics and dynamics of photoexcited quasi-particles involving excitons, which are valuable in aspects of application and understanding.

## Methods

**Synthesis and characterizations of the mixed $(F-PEA)_2(MA)_{n-1}Pb_nI_{3n+1}$ films.**
The mixed perovskite films were synthesized following a spin-coating process[45]. ITO-coated glass substrates were cleaned by ultrasonication for 20 min sequentially with detergent solution, deionized water, ethanol, and acetone, followed by drying in an $N_2$ stream then the treatment of ultraviolet-ozone for 30 min. The $(F-PEA)_2MA_3Pb_4I_{13}$ perovskite precursor solution was prepared by dissolving FACl of 8 mg, MAI of 119.2 mg, $PbI_2$ of 461 mg, and F-PEAI of 133.5 mg in a 1 mL mixed solvent of DMF and DMSO with a volume ratio of 95:5. The precursor solution with molar ratio corresponding to $(F-PEA)_2MA_3Pb_4I_{13}$ (n=4) was dropped on the preheated ITO substrate at 80 °C, and spin-coated at 6000 rpm for 70 s. Subsequently, the perovskite films were annealed at 100 °C for 10 min. The X-ray diffraction (Empyrean) examination reveals the high crystallinity of synthesized perovskites, as shown in

Supplementary Fig. S2. The thickness and surface morphology of sample were examined using scanning electron microscope (Nova NanoSEM 450). The optical absorption spectra were measured using UV−vis−NIR absorption spectroscopy (SolidSpec-3700). The fluorescence spectra were measured using fluorescence spectrophotometer (QuantaMaster 8000).

**Transient absorption measurement.**

The femtosecond TA measurements were performed on a Helios spectrometer (Ultrafast Systems) pumped by a Ti:sapphire regenerative amplifier (Legend Elite, Coherent) operating at 5 kHz with fundamental wavelength of 800 nm and pulse width of ∼40 fs. The excitation pulses were tuned in frequency by an optical parametric amplifier (Light Conversion). The probe beam with a wavelength range spanning over 420−800 nm was focused to a spot with a diameter of ∼30 μm, and the pump beam was focused to a spot with a diameter of ∼100 μm. The overall temporal resolution is estimated as ∼100 fs based on the cross-phase modulation signal measurement.

## References

1. Blancon, J.-C. et al. Scaling law for excitons in 2D perovskite quantum wells. *Nat. Commun.* **9**, 2254 (2018).
2. Tanaka, K. et al. Image charge effect on two-dimensional excitons in an inorganic-organic quantum-well crystal. *Phy. Rev. B* **71**, 045312 (2005).
3. Burgos-Caminal, A., Socie, E., Bouduban, M. E. F. & Moser, J. E. Exciton and Carrier Dynamics in Two-Dimensional Perovskites. *J. Phys. Chem. Lett.* **11**, 7692−7701 (2020).
4. Ponseca, C. S., Chábera, P., Uhlig, J., Persson, P. & Sundström, V. Ultrafast Electron Dynamics in Solar Energy Conversion. *Chem. Rev.* **117**, 10940−11024 (2017).
5. Fu, J., Ramesh, S., Melvin Lim, J. W. & Sum, T. C. Carriers, Quasi-particles, and Collective Excitations in Halide Perovskites. *Chem. Rev.* **123**, 8154−8231 (2023).
6  Kaindl, R. A., Hägele, D., Carnahan, M. A. & Chemla, D. S. Transient terahertz spectroscopy of excitons and unbound carriers in quasi-two-dimensional electron-hole gases. *Phys. Rev. B* **79**, 045320 (2009).
7  Sun, Q. et al. Ultrafast and High-Yield Polaronic Exciton Dissociation in Two-Dimensional Perovskites. *J. Am. Chem. Soc.* **143**, 19128−19136 (2021).

8    Xing, G. et al. Transcending the slow bimolecular recombination in lead-halide perovskites for electroluminescence. *Nat. Commun.* **8**, 14558 (2017).

9    Wang, N. et al. Perovskite light-emitting diodes based on solution-processed self-organized multiple quantum wells. *Nat. Photonics* **10**, 699−704 (2016).

10   Baranowski, M. & Plochocka, P. Excitons in metal-halide perovskites. *Adv. Energy Mater.* **10**, 1903659 (2020).

11   Mott, N. F. Metal-Insulator Transition. *Rev. Mod. Phys.* **40**, 677−683 (1968).

12   Suzuki, T. & Shimano, R. Exciton Mott Transition in Si Revealed by Terahertz Spectroscopy. *Phys. Rev. Lett.* **109**, 046402 (2012).

13   Palmieri, T. et al. Mahan excitons in room-temperature methylammonium lead bromide perovskites. *Nat. Commun.* **11**, 850 (2020).

14   Hansen, K. R. et al. Mechanistic origins of excitonic properties in 2D perovskites: Implications for exciton engineering. *Matter* **6**, 3463–3482 (2023).

15   Smith, I. C., Hoke, E. T., Solis-Ibarra, D., McGehee, M. D. & Karunadasa, H. I. A layered hybrid perovskite solar-cell absorber with enhanced moisture stability. *Angew. Chem. Int. Edit.* **53**, 11232−11235 (2014).

16   Liu, Y. et al. Ultrahydrophobic 3D/2D fluoroarene bilayer-based water-resistant perovskite solar cells with efficiencies exceeding 22%. *Sci. Adv.* **5**, eaaw2543 (2019).

17   Wygant, B. R., Ye, A. Z., Dolocan, A. & Mullins, C. B. Effects of Alkylammonium Choice on Stability and Performance of Quasi-2D Organolead Halide Perovskites. *J. Phys. Chem. C* **124**, 10887−10897 (2020).

18   Deng, S. et al. Long-range exciton transport and slow annihilation in two-dimensional hybrid perovskites. *Nat. Commun.* **11**, 664 (2020).

19   Shao, M. et al. Over 21% Efficiency Stable 2D Perovskite Solar Cells. *Adv. Mater.* **34**, 2107211 (2021).

20   Liu, J., Leng, J., Wu, K., Zhang, J. & Jin, S. Observation of Internal Photoinduced Electron and Hole Separation in Hybrid Two-Dimentional Perovskite Films. *J. Am. Chem. Soc.* **139**, 1432−1435 (2017).

21   Cao, D. H., Stoumpos, C. C., Farha, O. K., Hupp, J. T. & Kanatzidis, M. G. 2D Homologous Perovskites as Light-Absorbing Materials for Solar Cell Applications. *J. Am. Chem. Soc.* **137**, 7843−7850 (2015).

22   Hong, X. et al. Ultrafast charge transfer in atomically thin $MoS_2/WS_2$ heterostructures. *Nat. Nanotechnol.* **9**, 682−686 (2014).

23    Jiang, Y., Wei, J. & Yuan, M. Energy-Funneling Process in Quasi-2D Perovskite Light-Emitting Diodes. *J. Phys. Chem. Lett.* **12**, 2593−2606 (2021).

24    Fu, W., Chen, H. & Jen, A. K.-Y. Two-dimensional perovskites for photovoltaics. *Mater. Today Nano* **14**, 100117 (2021).

25    Marongiu, D., Saba, M., Quochi, F., Mura, A. & Bongiovanni, G. The role of excitons in 3D and 2D lead halide perovskites. *J. Mater. Chem. C* **7**, 12006−12018 (2019).

26    Righetto, M., Giovanni, D., Lim, S. S. & Sum, T. C. The photophysics of Ruddlesden-Popper perovskites: A tale of energy, charges, and spins. *Appl. Phys. Rev.* **8**, 011318 (2021).

27    Proppe, A. H. et al. Spectrally Resolved Ultrafast Exciton Transfer in Mixed Perovskite Quantum Wells. *J. Phys. Chem. Lett.* **10**, 419−426 (2019).

28    Zhang, T., Zhou, C., Lin, J. & Wang, J. Regulating the Auger Recombination Process in Two-Dimensional Sn-Based Halide Perovskites. *ACS Photonics* **9**, 1627−1637 (2022).

29    Ceballos, F., Cui, Q., Bellus, M. Z. & Zhao, H. Exciton formation in monolayer transition metal dichalcogenides. *Nanoscale* **8**, 11681−11688 (2016).

30    Chung, H. et al. Composition-Dependent Hot Carrier Relaxation Dynamics in Cesium Lead Halide ($CsPbX_3$, X=Br and I) Perovskite Nanocrystals. *Angew. Chem. Int. Edit.* **129**, 4224−4228 (2017).

31    Tao, W., Zhou, Q. & Zhu, H. Dynamic polaronic screening for anomalous exciton spin relaxation in two-dimensional lead halide perovskites. *Sci. Adv.* **6**, eabb7132 (2020).

32    Tao, W., Zhang, C., Zhou, Q., Zhao, Y. & Zhu, H. Momentarily trapped exciton polaron in two-dimensional lead halide perovskites. *Nat. Commun.* **12**, 1400 (2021).

33    Wu, X., Trinh, M. T. & Zhu, X.-Y. Excitonic many-body interactions in two-dimensional lead iodide perovskite quantum wells. *J. Phys. Chem. C* **119**, 14714−14721 (2015).

34    Yang, Y. et al. Observation of a hot-phonon bottleneck in lead-iodide perovskites. *Nat. Photonics* **10**, 53−59 (2015).

35    Oh, I.-K., Singh, J., Thilagam, A. & Vengurlekar, A. S. Exciton formation assisted by LO phonons in quantum wells. *Phys. Rev. B* **62**, 2045 (2000).

36    Siantidis, K., Axt, V. M. & Kuhn, T. Dynamics of exciton formation for near band-gap excitations. *Phys. Rev. B* **65**, 035303 (2001).

37  Luo, L. et al. Ultrafast terahertz snapshots of excitonic Rydberg states and electronic coherence in an organometal halide perovskite. *Nat. Commun.* **8**, 15565 (2017).

38  Liang, Y. et al. Solvent Recrystallization-Enabled Green Amplified Spontaneous Emissions with an Ultra-Low Threshold from Pinhole-Free Perovskite Films. *Adv. Funct. Mater.* **31**, 2106108 (2021).

39  Stern, M., Garmider, V., Umansky, V. & Bar-Joseph, I. Mott transition of excitons in coupled quantum wells. *Phys. Rev. Lett.* **100**, 256402 (2008).

40  Baldini, E., Palmieri, T., Dominguez, A., Rubio, A. & Chergui, M. Giant Exciton Mott Density in Anatase $TiO_2$. *Phys. Rev. Lett.* **125**, 116403 (2020).

41  Mahan, G. Excitons in metals. *Phys. Rev. Lett.* **18**, 12 (1967).

42  Sekiguchi, F. & Shimano, R. Excitonic correlation in the Mott crossover regime in Ge. *Phys. Rev. B* **91**, 155202 (2015).

43  Xing, G. et al. Long-range balanced electron-and hole-transport lengths in organic-inorganic $CH_3NH_3PbI_3$. *Science* **342**, 344-347 (2013).

44  Dou, L. et al. Atomically thin two-dimensional organic-inorganic hybrid perovskites. *Science* **349**, 1518-1521 (2015).

45  Shi, Jishan, et al. Fluorinated low dimensional Ruddlesden-Popper perovskite solar cells with over 17% power conversion efficiency and improved stability. *Adv.Mater.* **31**, 1901673 (2019).

## Acknowledgements

J.W., Z.L., and W.C. thank the supports from the Ministry of Science and Technology of China (2021YFB3800104), the National Natural Science Foundation of China (52002140, U20A20252), the Young Elite Scientists Sponsorship Program by CAST, the Natural Science Foundation of Hubei Province (2022CFA093), the Self-determined and Innovative Research Funds of HUST (2020kfyXJJS008). We thank the Analytical and Testing Center in Huazhong University of Science and Technology for the supports.

## Competing interests

The authors declare no competing interests.

.

# Supplementary information: Time-resolved optical assessment of exciton formation in mixed two-dimensional perovskite films

Zheng Zhang[1,2,3], Jianan Wang[1,3], Yijie Shi[1,2], Xi Wang[1,2], Zhong Wang[1,2], Xiangyu Zhu[1,2], Chunlong Hu[1,2], Zonghao Liu[1], Wei Chen[1], Wenxi Liang[1,2*]

## Supplementary figures

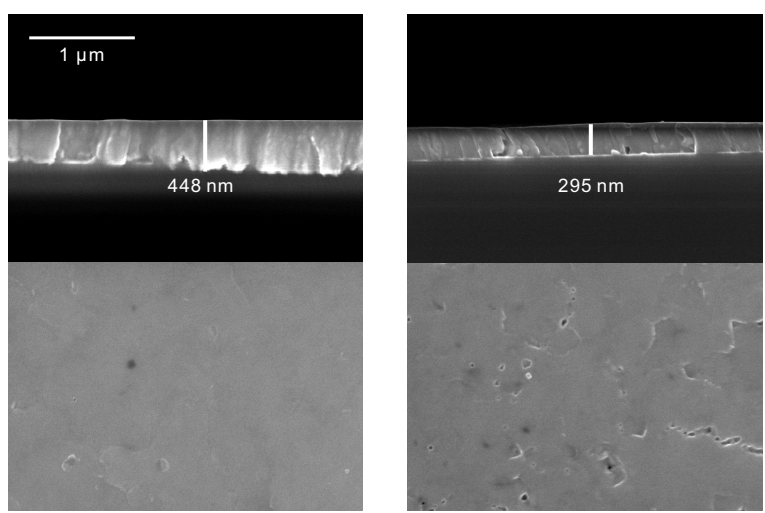

**Figure S1** Scanning electron microscope images of the synthesized thin films of mixed 2D (F-PEA)$_2$(MA)$_{n-1}$Pb$_n$I$_{3n+1}$ with different thicknesses. Left, cross-sectional (top) and surface (bottom) views of a sample with thickness of 448 nm. Right, cross-sectional (top) and surface (bottom) views of a sample with thickness of 295 nm.

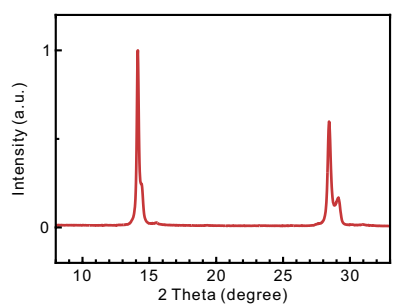

**Figure S2** X-ray diffraction pattern of a sample of mixed 2D (F-PEA)$_2$(MA)$_{n-1}$Pb$_n$I$_{3n+1}$.

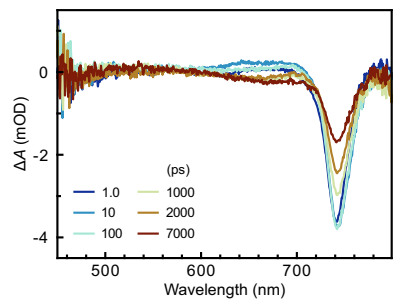

**Figure S3** Spectra of PB at selected delay times, under front excitation with 400 nm wavelength and fluence of 2.6 µJ/cm$^2$, extracted from Fig. 1b in the main text.

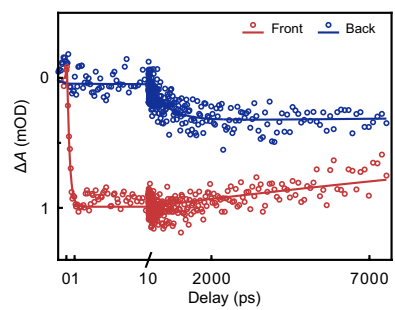

**Figure S4** Comparison of the kinetic traces of PB fitted with multiexponential function, under front and back excitations with 400 nm wavelength and fluences of 1.0 µJ/cm$^2$ and 2.6 µJ/cm$^2$, respectively. See Table S2 for the fitting parameters. The break on the x-axis represents the delay time of 10 ps.

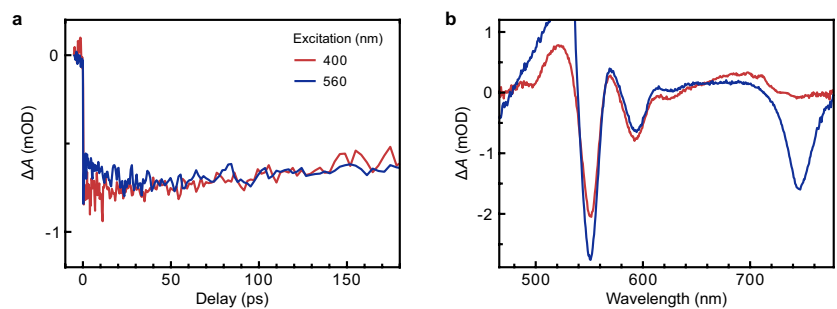

**Figure S5** Kinetic traces (**a**) of PB3 and spectra extracted at 1 ps (**b**) by back excitations with 400 nm, 2.6 µJ/cm$^2$, and 530 nm, 3.8 µJ/cm$^2$, respectively.

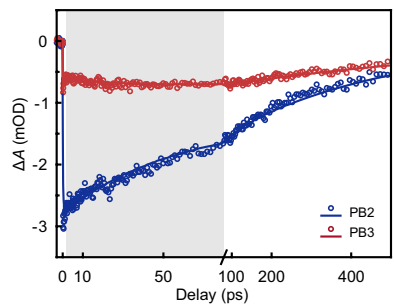

**Figure S6** Kinetic traces of PB2 and PB3 fitted with multiexponential function, by back excitation with 530 nm and fluence of 3.8 μJ/cm$^2$. The gray area highlights the duration of carrier transfer from n2 phase to n3 phase.

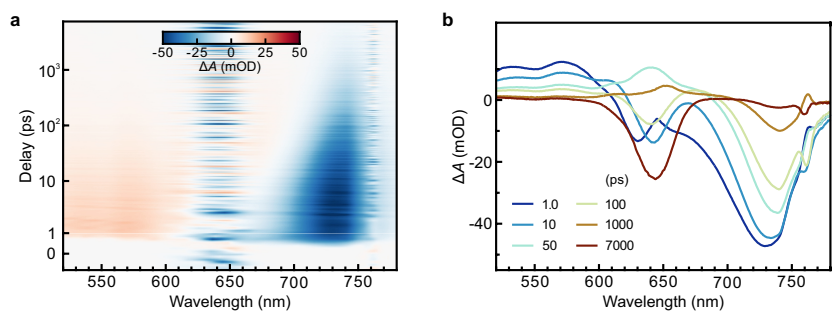

**Figure S7** Pseudocolor contour plot (**a**) and extracted spectra at selected delay times (**b**) of TA measurement by front excitation with 650 nm and fluence of 318.3 μJ/cm$^2$. Note that the noisy result of PB3 is due to the pump wavelengths close to the probed signals.

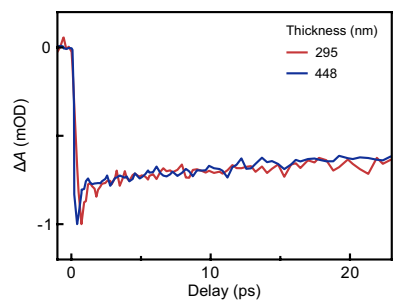

**Figure S8** Kinetic traces of PB2 by back excitation of 400 nm on samples with different thicknesses.

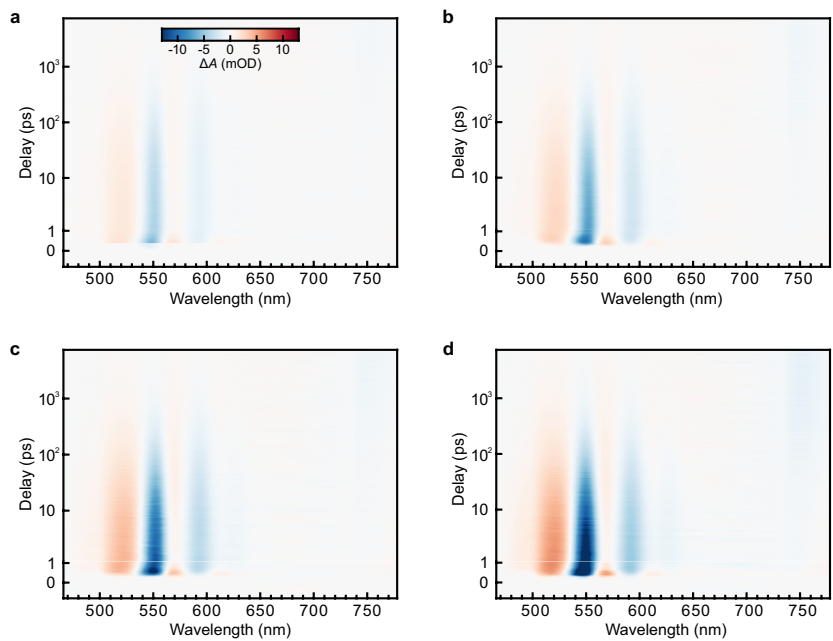

**Figure S9** Pseudocolor contour plots of TA measurement by back excitation with 400 nm and fluences of 2.6 (**a**), 5.1 (**b**), 7.7 (**c**), and 10.2 (**d**) μJ/cm$^2$, showing identical redshifts of PB2.

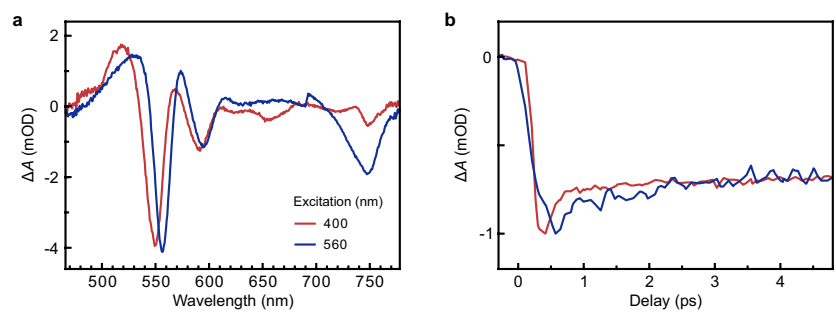

**Figure S10** Spectra extracted at 1 ps (**a**) and kinetic traces of PB2 (**b**) by back excitations with 400 and 560 nm, and fluences of ~10.0 μJ/cm$^2$.

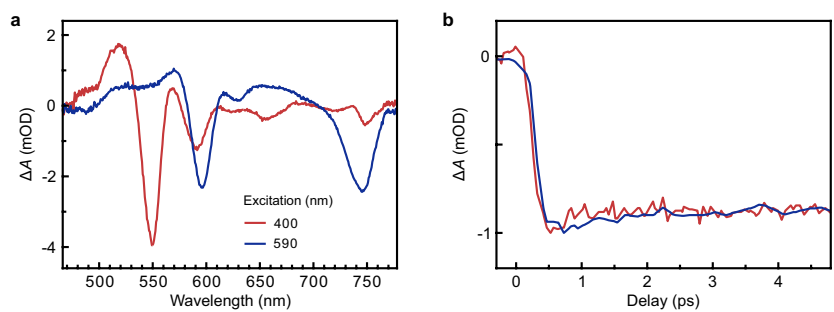

**Figure S11** Spectra extracted at 1 ps (**a**) and kinetic traces of PB3 (**b**) by back excitations with 400 and 590 nm, and fluences of ~10.0 μJ/cm$^2$.

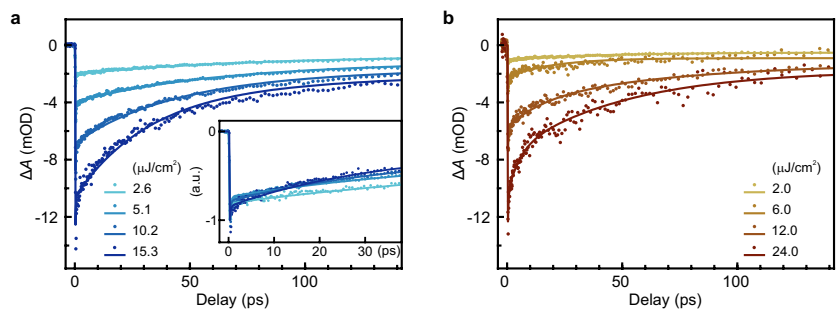

**Figure S12** Kinetic traces with multiexponential fits of PB2 excited with various fluences, by 400 nm (**a**) and 560 nm (**b**), respectively. Inset in **a**, normalized kinetic traces demonstrate the accelerating rates of exciton dissociation with increasing excitation fluences.

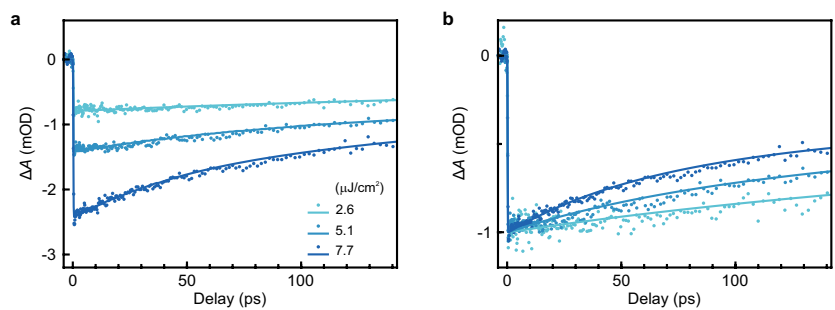

**Figure S13** Kinetic traces with multiexponential fits (**a**) of PB3 excited with 400 nm and various fluences, and normalized traces (**b**) demonstrating the accelerating rates.

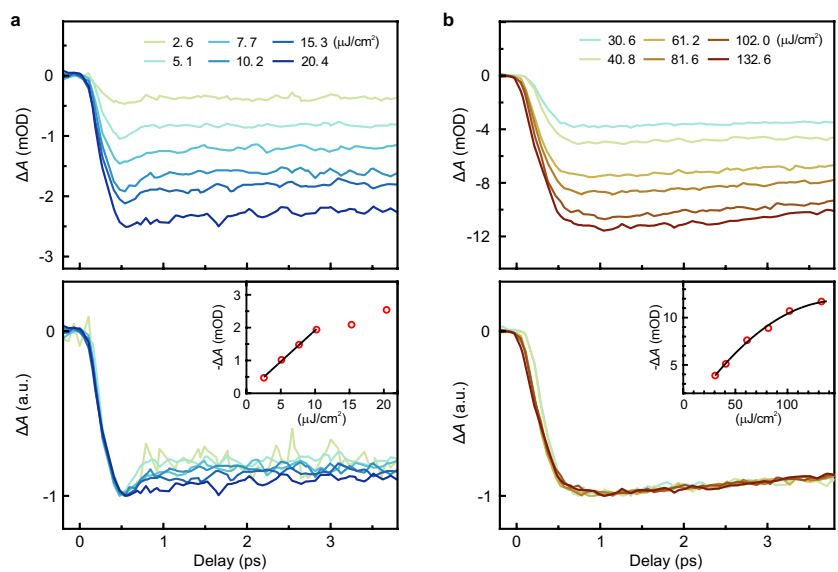

**Figure S14** Kinetic traces of PB3 by back excitation with 400 nm and various fluences. **a** and **b,** Kinetic traces (upper panels) and the normalized traces (lower panels) of PB3 excited with fluences of 2.6–20.4 µJ/cm$^2$ and 30.6–132.6 µJ/cm$^2$, respectively. Insets, linear (**a**) and saturating (**b**) trends of the PB3 amplitudes following the increase of excitation fluence. The black lines are the guides to the eye. It is noteworthy that the sub-picosecond decay of PB3 vanishes when the injected carrier density reaches 9.2×10$^{17}$ cm$^{-3}$ (excitation fluence of 20.4 µJ/cm$^2$), consistent with the picture that Mahan excitons exist in materials with large $E_b$, given that the $E_b$ of n3 phase is smaller than that of n2 phase, by our measurements and the results reported in literature. The kinetics of PB3 exhibit saturation under lower injected carrier densities compared to those of PB2, which can be attributed to the larger Bohr radius of excitons in n3 phase, resulting in a lower carrier density required for the stepping in of screening effect[1], or a lower Mott threshold for PB3, or the combination of both.

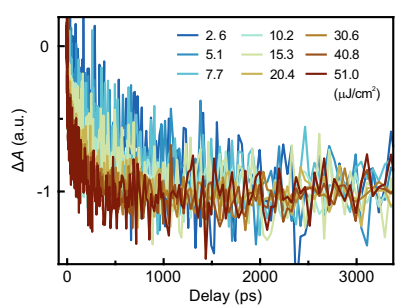

**Figure S15** Measured kinetic traces of PB signal following the increase of excitation fluence.

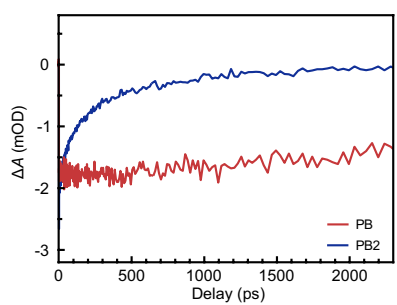

**Figure S16** Decays of PB by front excitation and PB2 by back excitation. The carriers in the 3D phase are mainly in the free state, while the carriers in the 2D phase are mainly in the form of exciton.

# Supplementary Discussions

**Supplementary Discussion 1:**

**Carrier transfers in different phases of the mixed (F-PEA)$_2$(MA)$_{n-1}$Pb$_n$I$_{3n+1}$**

The PB of back excitation shows barely visible signal in the early 100 ps, then a slow rise of ~600 ps followed by a process with barely decay, corresponding to a fitted lifetime of ~147 ns, see Fig. 1d and 1e in the main text, in sharp contrast to that of front excitation, which shows a rapid rise of ~0.3 ps followed by a long decay with fitted lifetime of ~32 ns, see Fig. 1b in the main text. Better comparison of two kinetic traces of PB is illustrated in Fig. S4, with fitting parameters tabulated in Table S2. The distinguishment between two traces testify that the 600 ps rise time by back excitation originates from the carrier transfer from the 2D phases to the 3D phase, while the 0.3 ps rise time by front excitation originates from the relaxation of hot carrier directly excited in the 3D phase[2]. The decays of PB2 and PB3 by back excitation are accompanied with the rise process of PB, as depicted in Fig. 1e in the main text, indicating the process of band filling in the 3D phase induced by the transferred carriers. Only PB but no PB2 or PB3 visible in the spectra by front excitation indicates that most of the excitation photons are absorbed by the 3D phase, hence the absence of signals from the 2D phases, neither the process of carrier transfer. We observe a factor of ~5 for the lifetime of PB by back excitation over that by front excitation, which may be attributed to the main contributor for the bleach signal as the transferred electrons from the 2D phases. Due to the band alignment of heterojunctions in the sample (see the next discussion), holes stay in the 2D phases, resulting in an electron-hole separation state, hence the longer lifetime of excited carriers.

The PB2 decays with a fitted lifetime of ~60 ps after the sup-picosecond decay process, while the trace of PB3 exhibits a plateau, as shown in Fig. 1e in the main text. The fitted result shows that this plateau rises in a duration of also ~60 ps. With the correlation of time span and opposite trends of kinetics, we infer that carrier transfer also takes place between the 2D phases. During the observed period of 60 ps, a portion of relaxed carriers residing at band edges in PB2 inject into PB3, cancelling out the decay of PB3 originated from the possible exciton-exciton annihilation due to strong quantum confinement[3], which are discussed in Supplementary Discussion 4. The scenario of carrier transfer between the 2D phases is corroborated by the measurements under varied excitation condition. We excited the sample with lower photon energy (530 nm) and higher fluence (3.8 μJ/cm$^2$) to reduce the proportion of excited carriers at band edge forming excitons, see discussions in the main text, and to increase the transferable carriers in n2 phase. The rise process of kinetics of PB3 due to the band filling by injected carriers is more prominent, as expected, as shown in Fig. S5 and S6.

The hot carriers generated by excitation of 530 nm possess lower initial energy and higher relaxation rate compared to those by excitation of 400 nm. As the relaxation of hot carriers takes place in a duration on sub-picosecond scale, way much faster than the observed transfer process between the 2D phases, we are able to exclude the impact of initial carrier temperature on the carrier transfer process. The injected carriers were kept in low density so that the results were measured in the linear absorption regime, see discussions in the main text, hence the amplitudes of PB2 and PB3 were proportional to the injected carrier density. The kinetic traces of PB3 under two excitation conditions are almost identical after 60 ps, as shown in Fig. S5a, indicating the same carrier populations in n3 phase after the early stage of several tens of picoseconds. A spectral comparison extracted at 1 ps, when the processes of hot carrier relaxation and exciton formation complete, see discussions in the main text, indicates that more carriers reside in n2 phase and less carriers in n3 phase in the case of excitation of 530 nm compared to that of 400 nm, as shown in Fig. S5b. The increased transferrable carriers between two phases results in a more prominent rise process with lifetime of ~40 ps in the kinetics of PB3, as shown in Fig. S6, with fitting parameters tabulated in Table S3.

**Supplementary Discussion 2:**
**Band alignment in the mixed $(F-PEA)_2(MA)_{n-1}Pb_nI_{3n+1}$**

We verify the band alignment among phases in the mixed 2D HOIPs to elucidate the process of carrier transfer, by the measurement of front excitation with pulses of 650 nm and fluence of 318.3 μJ/cm$^2$, which the photon energy is only enough to exclusively excite the 3D phase. A strong signal of PB accompanied with weak signals of PB2 and PB3 superimposing on the broadband absorption signal are observed, as shown in Fig. S7, indicating a process of carrier transfer from the 3D phase to the 2D phases, resulting in the band filling in the 2D phases, given the fact that the excitation of the phase with bands sitting on low positions in the energy diagram of heterostructure with staggered band alignment give rise to the hole transfer to the phase with bands on high positions. Combining the TA measurements by front and back excitations of 400 nm shown in Fig. 1 in the main text, we are confident to conclude a type-II band alignment for our samples, as illustrated in Fig. 1f.

**Supplementary Discussion 3:**
**Ruling out candidate processes for the sub-picosecond decays in PB2 and PB3**

We rule out the process of energy transfer by the following observations: (i) such ultrafast decay also takes place in 2D HOIPs with pure 2D phases[4,5]; (ii) the rate of energy transfer is correlated with the distance between donor and acceptor in heterojunctions[6], which is against the results of control measurements, showing that the sub-picosecond decay processes of PB2 on samples with thicknesses apparently different remained almost identical, as shown in Fig. S8. This independence rules out the process of hot carrier transfer as well, which is determined by the travelling distance[7]. We then rule out the defect associated capture process by the following observations: (i) the SEM inspections reveal that thicker the sample, better crystallized, thus less defects, so that the independence on sample thickness also indicates an independence on defect density; (ii) the organic cations providing protection to the inorganic octahedra is reported the reason of lower defect density for 2D perovskites compared to their 3D counterparts, but the observed sub-picosecond decay presents only in the 2D phases; (iii) the fact that mixed 2D perovskites are popular as efficient absorbers in solar cells[8] suggests an unlikely very high defect density, but the sub-picosecond decay persists when the injected carrier density is over the Mott criterions, see discussions in the main text; (iv) the process of sub-picosecond decay diminishes as the energy of excitation photon reducing from above the bandgap towards the bandgap, see discussions in the main text, which is not explained by the picture of defect capturing. The effects of bandgap renormalization, which increase with more injected carriers, are ruled out for corresponding the redshift during the sub-picosecond decay because the observed redshift is independent on excitation fluences, as shown in Fig. S9. Wu *et al.* reported a redshift in the bleach signal of $(C_4H_9NH_3I)_2(CH_3NH_3I)_{n-1}(PbI_2)_n$ on the same timescale with no dependence on injected carrier density, and concluded the formation of localized excitons from delocalized free carriers as the origin[9].

In our measurements of excitation energy dependence, we set the injected carrier densities in close values in the linear absorption regime, so that the variation of excitation condition was only the initial temperature of the excited carriers. Huang L. *et al.* found that carriers in perovskites excited with higher photon energy possess larger mobility and higher elastic speed[10], hence probably lower the possibility of being captured by defects. However, the observed rate of sub-picosecond decay goes faster under excitation with higher energy in our case, contradicting the picture of defect capturing. Surface defect capturing is neither likely the candidate accounting for the sub-picosecond decay, as shorter the excitation wavelength, shallower the penetration depth of excitation makes the bleach signal more sensitive to surface defects, hence the increasing of capture time[11], which is not the case of our observations, not even the vanish and reappearance of PB2 and PB3 under certain excitation wavelengths.

**Supplementary Discussion 4:**
**Dynamics associated to exciton dissociations**

The slow decay of PB2 by above-bandgap excitation (400 nm) increases in both rate and weight when the injected carrier density (i.e., the excitation fluence) increases, as shown in Fig. S12a the kinetics traces and multiexponential fits, with fitting parameters tabulated in Table S6. These trends can be interpreted with exciton-exciton annihilation[3], which is highly sensitive to the injected carrier density. The slow decays of PB2 by excitonic excitation (560 nm) show apparently shorter lifetimes, with the trends of increasing rates and decreasing weights, as shown in Fig. S12b with fitting parameters tabulated in Table S7. The slow decays of PB3 change in the similar trend but superimposed with an offset induced by the transferred carriers from n2 phase, as shown in Fig. S13, with fitting parameters tabulated in Table S8. The observed increases of decay rate of exciton dissociation may be attributed to the screening effects among carriers[12], which result in a temporal reduction of exciton binding energy, subsequently higher possibilities for excitons interacting with other quasiparticles, accelerating their dissociation into free carriers. Considering the slow decay processes are complicated by the concurrence of exciton dissociation, exciton-exciton annihilation, carrier transfers, and other possible processes associated with defects, here we address that the specific picture for the exciton dissociation goes beyond the scope of this work and is worthy further investigations.

**Supplementary Discussion 5:**
**Evaluation of $E_b$**

When $E_x$ is lower than the band gap, hot excitons, instead of hot carriers, are initially generated as the deposited photon energy is not enough to pump free carriers to the bottom of conduction band. Hot excitons release their extra energy to relax to the ground state of excitonic band, contributing to the bleach signal. At this point the sub-picosecond decay is absent in the bleach signal, as there is no process of free carriers forming into excitons. When $E_x$ is lower than the ground state of excitonic band $E_{1s}$, no more bleach signal is visible as there are neither free carriers nor bound excitons excited. The difference of $E_x$ for the measured TA features reflects the discrepancy of energy between the bottom of conduction band and the ground state of excitonic band. Such energy discrepancy is known as the exciton binding energy, $E_b$, so we have $E_b = E_g - E_{1s} = \frac{1240}{\lambda_1} - \frac{1240}{\lambda_{PB}}$. $\lambda_1$ is the excitation wavelength required for the vanishing of sub-picosecond decay, which is 500 nm for PB2. $\lambda_{PB}$ is the wavelength of probed bleaching peak during excitonic excitation, which is 557 nm for PB2. The calculated $E_b$ for n2 phase is 253 meV. For n3 phase, $\lambda_1$ is 560 nm and $\lambda_{PB}$ is 590 nm, yielding an $E_b$ of 113 meV. Considering the ~80 meV energy dispersion of the femtosecond pulses used in our TA measurements, the obtained $E_b$ values are satisfying in terms of accuracy.

**Supplementary Discussion 6:**
**Evaluation of free carrier and exciton occupations at quasi-thermal equilibrium**

The ratio of free carriers to excitons reaches quasi-thermal equilibrium within a few picoseconds after optical excitation. We can evaluate the occupation proportion free carriers and excitons at equilibrium using the Saha equation,

$$\frac{x^2}{1-x} = \frac{1}{n_0} \frac{(2\pi\mu k_B T)^{3/2}}{h^3} \exp\left(-\frac{E_b}{k_B T}\right)$$

where $x$ is the proportion of free carriers, $n_0$ is the injected carrier density, $\mu$ is the reduced mass of excitons setting to 0.2 times the electron mass[13], $k_B$ is the Boltzmann constant, $T$ is set to room temperature (300 K), $h$ is the Planck constant. Taking $1.15 \times 10^{17}$ cm$^{-3}$ of $n_0$ used in our experiment and the calculated $E_b$ of 249 and 133 meV for n2 and n3 phases, respectively, we obtain a proportion of 3% and 24% for free carriers in n2 and n3 phases, respectively, assuming all injected carriers by the excitation of 400 nm are populated in the free states.

**Supplementary Discussion 7:**
**Evaluation of contributions from free carriers and excitons to bleach signal**

Given the observed sub-picosecond duration of exciton formation and tens-of-picosecond duration of exciton dissociation, we assume that the process of exciton formation is complete and the interband relaxation process of exciton is not yet occurred at 1 ps. An injected carrier density of $1.15 \times 10^{17}$ cm$^{-3}$ in n2 phase yields 1.84 mOD for the maximum amplitude of bleach signal, assuming all injected carriers by the excitation of 400 nm are populated in the free states. The amplitude of bleach signal decreases to 1.40 mOD at 1 ps. Taking the calculated proportion of 3% for the left free carriers, we obtain 1.35 mOD for the contribution of formed excitons to the bleach signal at this delay time, hence the contribution of each free carrier to the amplitude of bleach signal is 1.32 times that of an exciton. By the same method, a factor of 1.25 is calculated for n3 phase.

**Supplementary Discussion 8:**

**Evaluation of Mott densities for n2 and n3 phases and calibration of injected carrier density**

We use the following expression to calculate the Mott density,

$$n_M = a_B^{-3} \frac{\pi 1.19^6}{4^3 \times 3} \left(\frac{m_e m_h}{(m_e + m_h)^2}\right)^3$$

where $a_B$ is the exciton Bohr radius, ~0.8 nm for n2 phase and ~1 nm for n3 phase[14], $m_e$ and $m_h$ is the effective mass of electrons and holes, respectively. Taking $m_e = 0.4\ m_0$ and $m_h = 0.6\ m_0$ for n2 phase[15], the Mott density is calculated as $1.25 \times 10^{18}$ cm$^{-3}$. Similarly, taking $m_e = 0.35\ m_0$ and $m_h = 0.58\ m_0$ for n3 phase, the Mott density is calculated as $6.01 \times 10^{17}$ cm$^{-3}$.

Assuming one absorbed photon generates one electron-hole pair, the density of photogenerated carriers can be estimated with the light intensity,

$$n = \alpha \frac{E\lambda}{hc \times \pi r^2 d}(1 - R_{pump} - T_{pump})$$

where α=1 is the proportion of electron-hole pairs generated per absorbed photon, $E$ is the energy of a single pulse at a certain wavelength $\lambda$, $h$ is the Planck constant, $c$ is the speed of light, $r$ is the spot radius of excitation, ~100 μm in our measurements, $d$ is the sample thickness, 448 nm in our case. $R_{pump}$ and $T_{pump}$ are the reflectance and transmittance, respectively, measured for each used excitation wavelength. To inject carrier of the calculated Mott density of $1.25 \times 10^{18}$ cm$^{-3}$ and $6.01 \times 10^{17}$ cm$^{-3}$ for n2 and n3 phase, the required excitation fluences are 25.5 μJ/cm$^2$ and 12.3 μJ/cm$^2$, respectively.

**Supplementary Tables**

**Table S1** Fitting parameters of the kinetic traces shown in Fig. 1e in the main text.

| Peak | $A_1$ | $\tau_1$ (ps) | $A_2$ | $\tau_2$ (ps) | $A_3$ | $\tau_3$ (ps) |
|---|---|---|---|---|---|---|
| PB2 | -0.316 | 0.2 | -0.324 | 59.9 | -0.360 | 604.0 |
| PB3 | 1 | 59.0 | -0.741 | 67.1 | -0.259 | 917.0 |
| PB | 1 | 599.0 | -1 | 147000.0 | | |

**Table S2** Fitting parameters of the kinetic traces of PB shown in Fig. S4.

| Excitation direction | $A_1$ | $\tau_1$ (ps) | $A_2$ | $\tau_2$ (ns) |
|---|---|---|---|---|
| Back | 1 | 599.0 | -1 | 147.0 |
| Front | 1 | 0.3 | -1 | 32.0 |

**Table S3** Fitting parameters of the kinetic traces of PB2 and PB3 shown in Fig. S6.

| Peak | $A_1$ | $\tau_1$ (ps) | $A_2$ | $\tau_2$ (ps) | $A_3$ | $\tau_3$ (ps) |
|---|---|---|---|---|---|---|
| PB2 | -0.161 | 0.3 | -0.273 | 40.0 | -0.565 | 410.0 |
| PB3 | 1 | 40.0 | -0.741 | 67.1 | -0.269 | 691.0 |

**Table S4** Fitting parameters of all measured kinetic traces of PB2.

| Wavelength (nm) | A1 | τ1 (ps) | A2 | τ2 (ps) |
|---|---|---|---|---|
| 400 | -0.65 | 0.18 | -0.35 | 60.0 |
| 420 | -0.62 | 0.30 | -0.38 | 61.3 |
| 440 | -0.33 | 0.33 | -0.67 | 64.4 |
| 460 | -0.32 | 0.36 | -0.68 | 55.0 |
| 480 | -0.28 | 0.42 | -0.72 | 58.6 |
| 500 | 0 | 0 | -1 | 58.0 |
| 520 | 0 | 0 | -1 | 63.8 |
| 530 | 0 | 0 | -1 | 64.0 |
| 535 | 0 | 0 | -1 | 48.0 |
| 550 | -0.48 | 0.35 | -0.52 | 35.2 |
| 560 | -0.59 | 0.35 | -0.41 | 20.0 |

**Table S5** Fitting parameters of all measured kinetic traces of PB3.

| Wavelength (nm) | A1 | τ1 (ps) | A2 | τ2 (ps) |
|---|---|---|---|---|
| 400 | -0.31 | 0.18 | -0.69 | 95 |
| 420 | -0.28 | 0.25 | -0.72 | 88 |
| 440 | -0.24 | 0.27 | -0.76 | 85 |
| 460 | -0.22 | 0.28 | -0.78 | 86 |
| 480 | -0.18 | 0.28 | -0.72 | 75 |
| 500 | -0.17 | 0.30 | -0.83 | 90 |
| 520 | -0.11 | 0.32 | -0.89 | 75 |
| 530 | -0.06 | 0.35 | -0.94 | 82 |
| 535 | -0.05 | 0.42 | -0.95 | 90 |
| 550 | -0.02 | 0.45 | -0.98 | 93 |
| 560 | 0 | 0 | -1 | 90 |
| 570 | 0 | 0 | -1 | 89 |
| 580 | 0 | 0 | -1 | 81 |
| 590 | -0.11 | 1.11 | -0.89 | 90 |

**Table S6** Fitting parameters of the kinetic traces of PB2 shown in Fig. S12a.

| Fluences ($\mu J/cm^2$) | A1 | $\tau_1$ (ps) | A2 | $\tau_2$ (ps) | A3 | $\tau_3$ (ps) |
|---|---|---|---|---|---|---|
| 2.6 | -0.552 | 0.20 | -0.209 | 60.00 | -0.239 | Inf |
| 5.1 | -0.663 | 0.20 | -0.213 | 53.80 | -0.114 | Inf |
| 10.2 | -0.538 | 0.20 | -0.304 | 41.80 | -0.158 | Inf |
| 15.3 | -0.449 | 0.20 | -0.392 | 32.60 | -0.159 | inf |

**Table S7** Fitting parameters of the kinetic traces of PB2 shown in Fig. S12b.

| Fluences ($\mu J/cm^2$) | A1 | $\tau_1$ (ps) | A2 | $\tau_2$ (ps) | A3 | $\tau_3$ (ps) |
|---|---|---|---|---|---|---|
| 2.6 | -0.359 | 0.35 | -0.322 | 40.00 | -0.319 | Inf |
| 7.7 | -0.395 | 0.35 | -0.345 | 20.00 | -0.260 | Inf |
| 15.3 | -0.400 | 0.35 | -0.204 | 15.00 | -0.396 | Inf |
| 30.6 | -0.272 | 0.35 | -0.244 | 5.00 | -0.484 | inf |

**Table S8** Fitting parameters of the kinetic traces of PB3 shown in Fig. S13.

| fluences ($\mu J/cm^2$) | $A_1$ | $\tau_1$ (ps) | $A_2$ | $\tau_2$ (ps) |
|---|---|---|---|---|
| 2.6 | -0.087 | 100 | -0.913 | 808 |
| 5.1 | -0.261 | 78 | -0.749 | 750 |
| 7.7 | -0.411 | 60 | -0.589 | 713 |

**Table S9** Fitting parameters of the kinetic traces of PB2 shown in Fig. 4a and 4b in the main text.

| Fluences (μJ/cm$^2$) | A1 | τ1 (ps) | A2 | τ2 (ps) |
|---|---|---|---|---|
| 2.6 | -0.698 | 0.15 | -0.302 | 60.0 |
| 5.1 | -0.724 | 0.15 | -0.276 | 60.0 |
| 7.7 | -0.747 | 0.15 | -0.253 | 58.6 |
| 10.2 | -0.727 | 0.16 | -0.273 | 57.0 |
| 15.3 | -0.642 | 0.16 | -0.358 | 41.3 |
| 20.4 | -0.771 | 0.15 | -0.329 | 30.0 |
| 30.6 | -0.663 | 0.17 | -0.337 | 23.6 |
| 40.8 | -0.649 | 0.17 | -0.354 | 21.9 |
| 61.2 | -0.538 | 0.18 | -0.462 | 20.0 |
| 81.6 | -0.429 | 0.20 | -0.571 | 16.4 |
| 102.0 | -0.348 | 0.22 | -0.652 | 14.7 |
| 132.6 | -0.241 | 0.3 | -0.769 | 14.4 |

**Table S10** Fitting parameters of the kinetic traces of PB shown in Fig. 4d in the main text.

| fluences (μJ/cm$^2$) | A1 | τ1 (ps) | A2 | τ2 (ps) |
|---|---|---|---|---|
| 2.6 | 1 | 599 | -1 | inf |
| 5.1 | 1 | 620 | -1 | inf |
| 7.7 | 1 | 596 | -1 | inf |
| 10.2 | 1 | 520 | -1 | inf |
| 15.3 | 1 | 368 | -1 | inf |
| 20.4 | 1 | 169 | -1 | inf |
| 30.6 | 1 | 128 | -1 | inf |
| 40.8 | 1 | 97 | -1 | inf |
| 51.0 | 1 | 78 | -1 | inf |

# References

1. Baldini, E., Palmieri, T., Dominguez, A., Rubio, A. & Chergui, M. Giant Exciton Mott Density in Anatase $TiO_2$. *Phys. Rev. Lett.* **125**, 116403 (2020).

2. Wang, L. et al. Observation of a phonon bottleneck in copper-doped colloidal quantum dots. *Nat. Commun.* **10**, 4532 (2019).

3. Yin, Z. et al. Auger-Assisted Electron Transfer between Adjacent Quantum Wells in Two-Dimensional Layered Perovskites. *J. Am. Chem. Soc.* **143**, 4725−4731 (2021).

4. Zhang, T., Zhou, C., Lin, J. & Wang, J. Regulating the Auger Recombination Process in Two-Dimensional Sn-Based Halide Perovskites. *ACS Photonics* **9**, 1627−1637 (2022).

5. Qin, C. et al. Carrier dynamics in two-dimensional perovskites: Dion–Jacobson vs. Ruddlesden–Popper thin films. *J. Mater. Chem. A* **10**, 3069−3076 (2022).

6. DuBose, J. T. & Kamat, P. V. Directing Energy Transfer in Halide Perovskite-Chromophore Hybrid Assemblies. *J. Am. Chem. Soc.* **143**, 19214−19223 (2021).

7. Li, M. et al. Slow cooling and highly efficient extraction of hot carriers in colloidal perovskite nanocrystals. *Nat. Commun.* **8**, 14350 (2017).

8. Shao, M. et al. Over 21% Efficiency Stable 2D Perovskite Solar Cells. *Adv. Mater.* **34**, 2107211 (2021).

9. Wu, X., Trinh, M. T. & Zhu, X.-Y. Excitonic many-body interactions in two-dimensional lead iodide perovskite quantum wells. *J. Phys. Chem. C* **119**, 14714−14721 (2015).

10. Guo, Z. et al. Long-range hot-carrier transport in hybrid perovskites visualized by ultrafast microscopy. *Science* **356**, 59−62 (2017).

11. Liu, H. et al. Ultrafast Anisotropic Evolution of Photoconductivity in $Sb_2Se_3$ Single Crystals. *J. Phys. Chem. Lett.* **13**, 4988−4994 (2022).

12. Sekiguchi, F. & Shimano, R. Excitonic correlation in the Mott crossover regime in Ge. *Phys. Rev. B* **91**, 155202 (2015).

13. Sun, Q. et al. Ultrafast and High-Yield Polaronic Exciton Dissociation in Two-Dimensional Perovskites. *J. Am. Chem. Soc.* **143**, 19128−19136 (2021).

14. Blancon, J. C. et al. Scaling law for excitons in 2D perovskite quantum wells. *Nat. Commun.* **9**, 2254 (2018).

15. Zhao, Y.-Q. et al. Layer-dependent transport and optoelectronic property in two-dimensional perovskite: $(PEA)_2PbI_4$. *Nanoscale* **10**, 8677−8688 (2018).